\documentclass[epjST,final]{svjour}
\usepackage{graphicx}
\usepackage{xcolor}
\usepackage{amsmath}
\begin{document}
\title{Characterizing multistability regions
in the parameter space of the Mackey-Glass delayed system}
\author{Juan P. Tarigo \and Cecilia Stari \and Cecilia Cabeza \and
  Arturo C. Marti}
\institute{Instituto de Física, Universidad de la República,
  Montevideo, Uruguay}

\abstract{Proposed to study the dynamics of physiological systems in
  which the evolution depends on the state in a previous time, the
  Mackey-Glass model exhibits a rich variety of behaviors including
  periodic or chaotic solutions in vast regions of the parameter
  space. This model can be represented by a dynamical system with a
  single variable obeying a delayed differential equation. Since it is
  infinite dimensional requires to specify a real function in a finite
  interval as an initial condition. Here, the dynamics of the
  Mackey-Glass model is investigated numerically using a scheme
  previously validated with experimental results. First, we explore
  the parameter space and describe regions in which solutions of
  different periodic or chaotic behaviors exist. Next, we show that
  the system presents regions of multistability, i.e.  the coexistence
  of different solutions for the same parameter values but for
  different initial conditions. We remark the coexistence of periodic
  solutions with the same period but consisting of several maximums
  with the same amplitudes but in different orders. We
    characterize the multistability regions by introducing families of
    representative initial condition functions and evaluating the
    abundance of the coexisting solutions. These findings contribute
  to describe the complexity of this system and explore the
  possibility of possible applications such as to store or to code
  digital information.  } 
\maketitle
\section{Introduction}
\label{intro}
The Mackey-Glass (MG) model was first introduced to model respiratory
and hemato\-poietic diseases related to physiological systems
\cite{mackey1977oscillation}.  Perhaps the most remarkable
characteristic of this model is that the evolution of the system
depends on the state in a previous, or \textit{delayed}, time.  MG
model obtained great recognition thanks to its ability to accurately
describe in simple terms complicated dynamics such as a variety of
human illnesses \cite{belair1995dynamical}. Nonetheless, the relevance
of the MG model goes beyond its application to specific systems and
results illuminating in a broad variety of delayed systems
\cite{biswas2018time} exhibiting chaotic behaviour and multistability.

Delayed systems are in general considerably more complicated than
non-delayed. Even in the simplest case of a single delay, $tau$, the
evolution of the system at present time $t$ depends on the state at
time $t- \tau$, thus, it also depends on a infinite set of previous
times, $t-\tau$, $t - 2 \tau$, .... These systems are infinite
dimensional and mathematically they can be represented by systems of
delayed differential equations (DDE). In this sense these equations
are far more complex than ODEs and behave like infinite-dimensional
systems of ODEs \cite{hale2013introduction}. Moreover, the dynamics of
  DDEs are far more rich than those of ODEs. For instance, the
  peculiar routes to chaos, or the creation and destruction of
  isolated peaks in the MG system have been detailed studied by Junges
  and Gallas \cite{junges2012intricate}.

The presence of multistability in the MG model, i.e. several
coexisting solutions for the same parameter values but different
initial conditions, was reported by \cite{losson1993solution}.  This
coexistence of solutions of a time-delayed feedback system could be of
practical interest. In particular, it was proposed as an alternative
way for storing information
\cite{mensour1995controlling,lim1998experimental,zhou2007isochronal}. The
ability to synchronize several coupled MG systems is also relevant and
has received considerable attention in the last years
\cite{pyragas1998synchronization,kim2006synchronization,shahverdiev2006chaos}.

In addition to numerical studies, different experimental systems based
on electronic implementations that mimic the MG model were proposed
\cite{namajunas1995electronic,amil2015exact}. Due to the analytical
and numerical difficulties, experimental results play a major role and
contribute to validate theoretical models. In particular, experimental
observations can shed light on the feasibility of the observed
solutions under the presence of noise or parameter mismatch.
Recently, an electronic system mimicking the dynamics of the MG
system, whose central elements are a bucket bridge device (BBD) and a
nonlinear function block, was proposed by Amil \textit{et al.}
\cite{amil2015exact}. A remarkable characteristic of this approach is
that the temporal integration is exact, thus, experimental and
numerical simulations agree very well with each other enabling the
study of large regions in the parameter space. In a subsequent
investigation \cite{amil2015organization}, this approach was used to
explore the parameter space described in terms of the dimensionless
delay and production rate. This study unmasked the existence of
periodic and chaotic solutions intermingled in vast regions of the
parameter space. A remarkable point is the existence of abundant
solutions with the same period but consisting of several peaks with
the same amplitudes but in different alternations.  These periodic
solutions can be translated to sequences of letters and classified
using symbolic algorithms.

In spite of these extensive studies, there are still numerous open
questions, in particular, in which regions of the parameter space the
system presents multistability and where the coexistence of solutions
is more abundant.  Here we report the presence of tongue-like
structures of stability in the midst of chaotic regions in the
parameter space and that these solutions seem to retain some of their
structure inside these regions of stability.  We also go deeper into
the organization of the solutions and the impact of
multistability. Given that there are infinite initial condition
functions, we select a specific family of functions as a proxy to
quantitatively evaluate the \textit{strength} of the multistabilty.
Our study shows that, although there are some general trends, there
exist regions in which the coexistence is clearly more noticeable than
in others. The rest of this work is organized as follows. In the next
Section we shortly summarize the basic ingredient of the model and
review the numerical methods used. In Section 3 we numerically explore
the parameter space identifying the most interesting regions. The
analysis of the multistability and the abundance of coexistent
solutions with the same parameter values but different initial
conditions is presented in Section 4. Finally, Section 5 is devoted to
the final remarks.

\section{The MG model and the numerical method}

 The original model describes the dynamics of a physiological
 variable, $P(t)$ representing the concentration of a particular cell
 population in the blood. The temporal evolution is governed by the
 following equation
\begin{equation}
    \frac{dP}{dt}=\frac{\beta_0 \theta^n P_\tau}{\theta^n+P_\tau^n}-\gamma P
    \label{eqMG}
\end{equation}
where $P_\tau=P(t-\tau)$ is the delayed variable and $\beta_0$,
$\theta$, $\tau$, $n$, $\gamma$ are real parameters
\cite{mackey1977oscillation}.  In this equation the first term in the
right hand side represents the nonlinear, delayed, production and the
second term accounts for the natural decaying.  The number of
parameters can be reduced by re-scaling the variables $x=P/\theta$ and
$t'=\gamma t$ in Eq. \eqref{eqMG} obtaining,
\begin{equation}
    \frac{dx}{dt'}=\alpha \frac{x_\Gamma}{1+x_\Gamma^n} - x
    \label{eqDimensionless}
\end{equation}
where $x_{\Gamma}=x(t'-\Gamma)$, $\alpha=\beta_0 / \gamma$ and
$\Gamma=\gamma \tau$. From now on we denote the re-scaled time simply
as $t$.

The temporal evolution of the dimensionless variable $x(t)$ is given
by Eq. \eqref{eqDimensionless} which is the center of our
analysis. This equation depends on three parameters: $n$, $\alpha$,
and $\Gamma$. The first one, $n$ is directly related to the mechanism
of production of the particular blood component and it is kept fixed
at $n=4$ through all this work. The other two parameters, $\alpha$ and
$\Gamma$, correspond to the production rate and the decay and we will
take them as the main variables of the control or parameter space.  In
this DDE with a single delay, the initial condition must specify a
function in the interval $(-\Gamma,0)$ to univocally determine the
solution $x(t)$.

The numerical integration of the DDE requires to redesign a standard
numerical methods for ordinary differential equations
\cite{smith2011introduction}. The first alternative is to appeal to
standard numerical methods, like Runge-Kutta schemes with constant
step-size, and store at each step the previous values of the variables
in an interval at least equal to the maximum delay. Moreover, in the
first steps during a time lapse at least equal to the maximum delay it
is necessary to rely on another method to advance. In general, this
approach is not the most efficient and it is difficult to verify the
stability and accuracy. Other methods to solve DDEs
  numerically include the Bellman's method of steps and wave
  relaxation methods as shown in \cite{bellen2013numerical}.

In a very different approach \cite{amil2015exact}, we proposed an
electronic system that reproduces MG model. Moreover, we showed that
taking into account the specific expressions for the nonlinear
function and the delay block an explicit discretization scheme can be
derived. This scheme presents the advantage that in addition to being
more natural and easier to implement than standard methods the
temporal evolution is exact in time.

This exact discretization of the Mackey-Glass equation \cite{amil2015exact} can be obtained by taking discrete times $t_j= j dt$, where $dt$ is the sampling time.
The number of values stored in the BBD,  $N$ (in this work, N=1194),  the time delay, $\Gamma$
and the sampling time are related through  $dt=\Gamma/ N $. The MG variable $x(t)$ is also discretized as $x_j$ whose temporal evolution is governed by the following equation
\begin{equation}
    x_{j+1} = x_j e^{-\frac{\Gamma}{N}} + \Big(1-e^{-\frac{\Gamma}{N}}\Big)\frac{\alpha x_{j-N+1}}{1+x_{j-N+1}}.
    \label{eqDiscrete}
\end{equation}

It can be shown that this numerical method based on the exact
discretization is a reliable representation of the experimental
electronic system and a good approximation of
Eq. \eqref{eqDimensionless} for sufficiently large values of $N$, as
well as being more efficient than standard numerical algorithms
\cite{amil2015exact,amil2015organization}.

\section{Exploring the parameter space}
\label{parameterSpace}

\begin{figure}
\centerline{\includegraphics[width=\columnwidth]{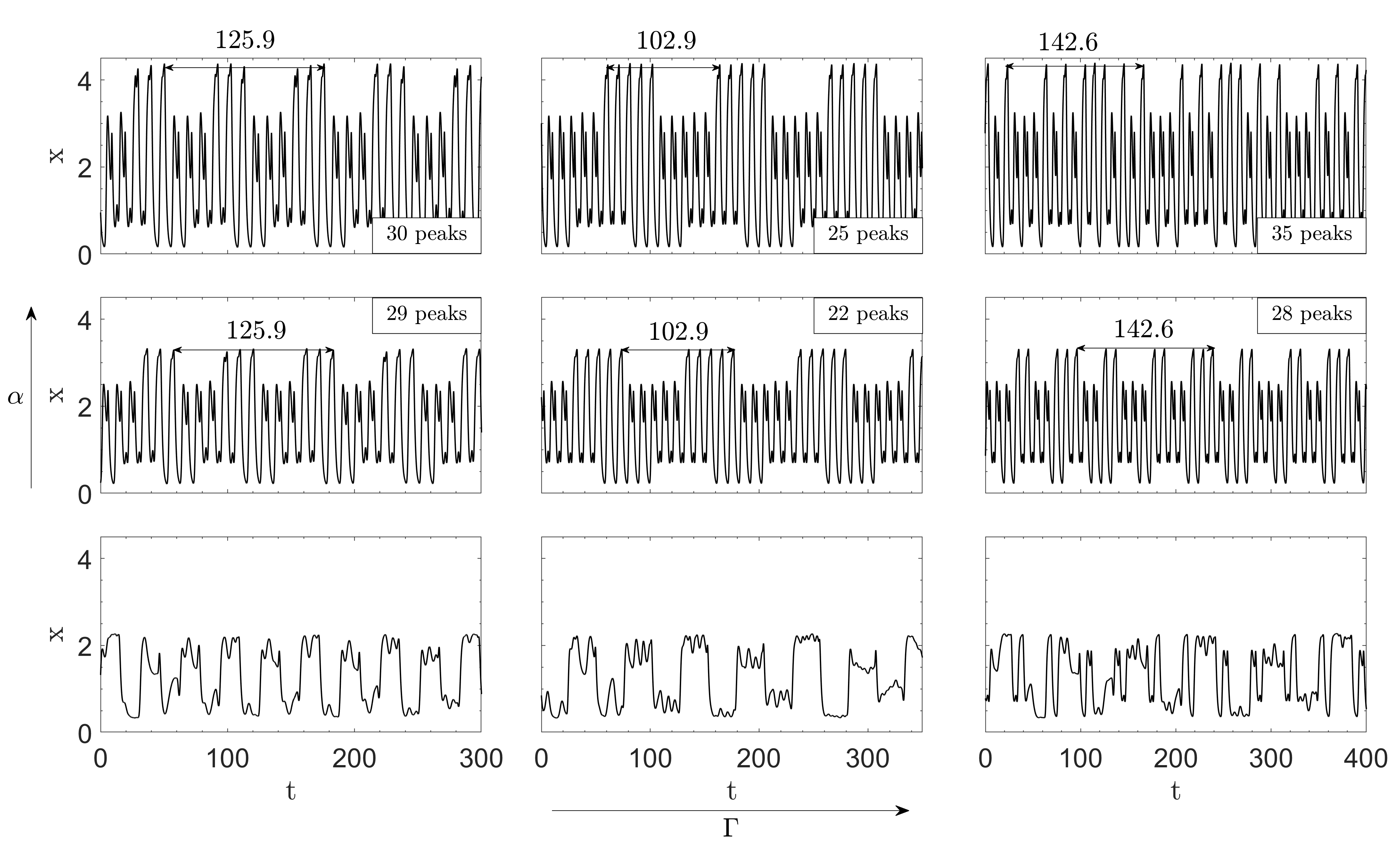}}
\caption{Time series evolution as control parameters $\alpha$ and
  $\Gamma$ are varied. Columns correspond to $\Gamma = [15, 25, 35]$
  from left to right and rows correspond to $\alpha = [4, 6, 8]$ from
  bottom to top. A transient time of $5000 \Gamma$ was neglected. In
  all the panels, initial conditions were set to $x_{in}(t)=1$
  $\forall t\in(-\Gamma,0)$. }
\label{figTimeSeries}
\end{figure}

As mentioned, the Mackey-Glass system exhibits a rich variety of
dynamics depending on the parameters: the nonlinear function exponent
$n$, the production rate, $\alpha$, and the delay $\Gamma$, and also
on the initial condition function. We start considering temporal
series for representative values of the delay, $\Gamma=15, 25, 35$ and
$\alpha=4, 6, 8$ while the initial condition function and the exponent
$n$ are kept fixed.  In Fig.~\ref{figTimeSeries} the temporal series
shown exhibit periodic behaviour for $\alpha = 6$ and $\alpha = 8$ for
all $\Gamma$ values. The number of peaks per period is 30, 25 and 35
for the top series and 29, 22 and 28 for the ones in the middle. The
bottom series, corresponding to $\alpha = 4$, are all chaotic.  We
observe that series in the same row reveal different period while
their shape changes abruptly. In this case, $T=125.9$, $T=102.9$ and
$T=142.9$, for $\Gamma=15$, $\Gamma=25$ and $\Gamma=35$ respectively.
The creation and destruction of peaks manifested in series along the
same column is clear where the temporal series present the same period
but different peaks count (indicated in each panel).

\begin{figure}
\centerline{\includegraphics[width=\columnwidth]{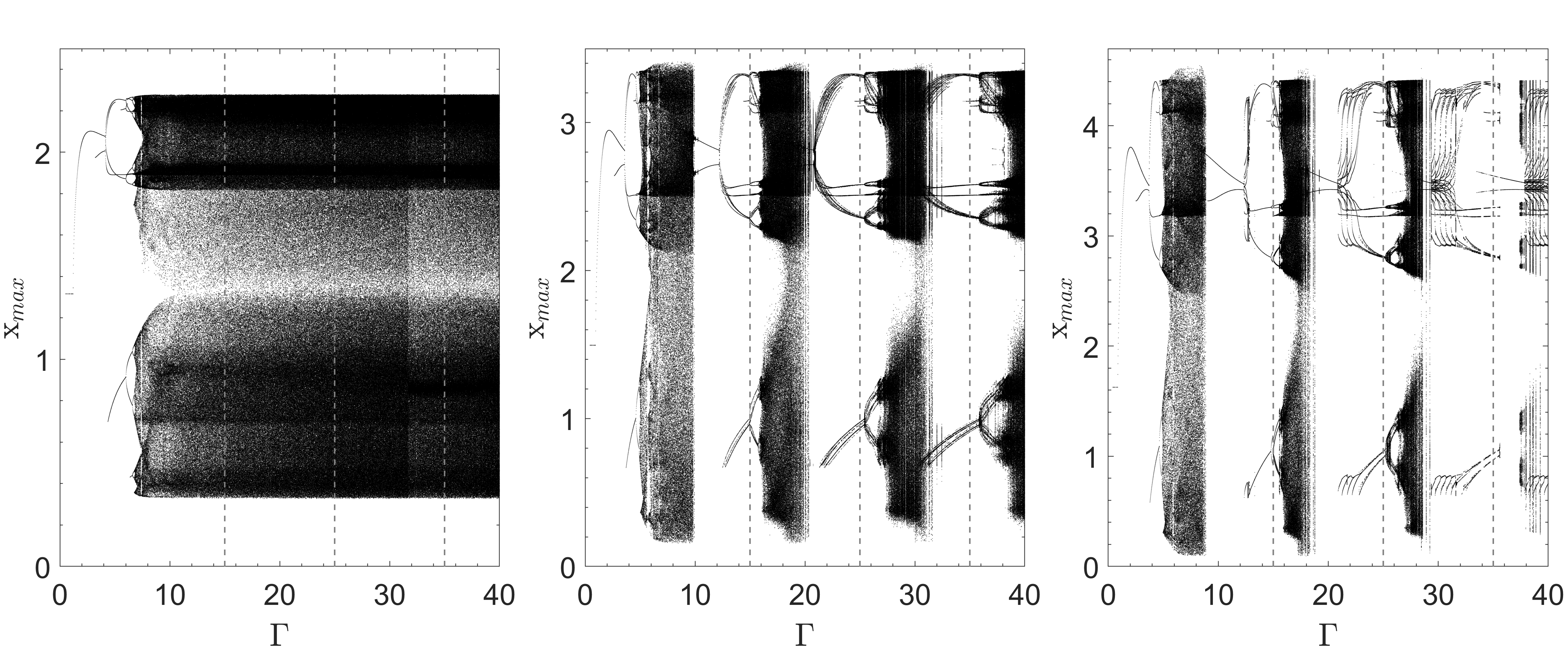}}
\caption{Bifurcation diagrams for different values $\alpha = 4$ (left
  panel), $\alpha = 6$ (center panel) and $\alpha = 8$ (right
  panel). Peaks are recorded in a span of $200 \Gamma$ after a
  transient of $5000 \Gamma$ was neglected. Initial conditions were
  set to $x_{in}(t)=1$ $\forall t\in(-\Gamma,0)$. Vertical dashed
  lines correspond to the parameter values of
  Fig.~\ref{figTimeSeries}. }
\label{figBifurcations}
\end{figure}

Figure \ref{figBifurcations} shows bifurcation diagrams obtained by plotting the local maximums (peaks) of the numerically computed solutions as a function of $\Gamma$ for three different values of  $\alpha$. In all the cases, the initial condition functions is  $x_{in}(t)=1$ in the interval $(-\Gamma, 0)$. These diagrams show the familiar periodic branches and chaotic behaviour. For $\alpha = 4$ and $\Gamma<8$, typical period-doubling branches are exhibited, and the system presents a no return to periodicity for $\Gamma >8$. For $\alpha = 6$ and $\alpha = 8$ three different periodic regions are observed after the first transition to chaos. Creation and destruction of branches is also evident in Fig.~\ref{figBifurcations} for all three diagrams, which is typical  of delayed systems. All periodic regions appear similar inside the diagrams even as the delay parameter $\Gamma$ is increased. The diagrams also  reveals regions of periodic solutions with similar peak structures
for a range of values of $\alpha$. In general, complex arrangements appear for high $\Gamma$ values as $\alpha$ is increased. For example the creation of a high number of spontaneous branches is shown for $\Gamma=30$ and $\alpha = 8$. 

\begin{figure}
\centerline{\includegraphics[width=\columnwidth]{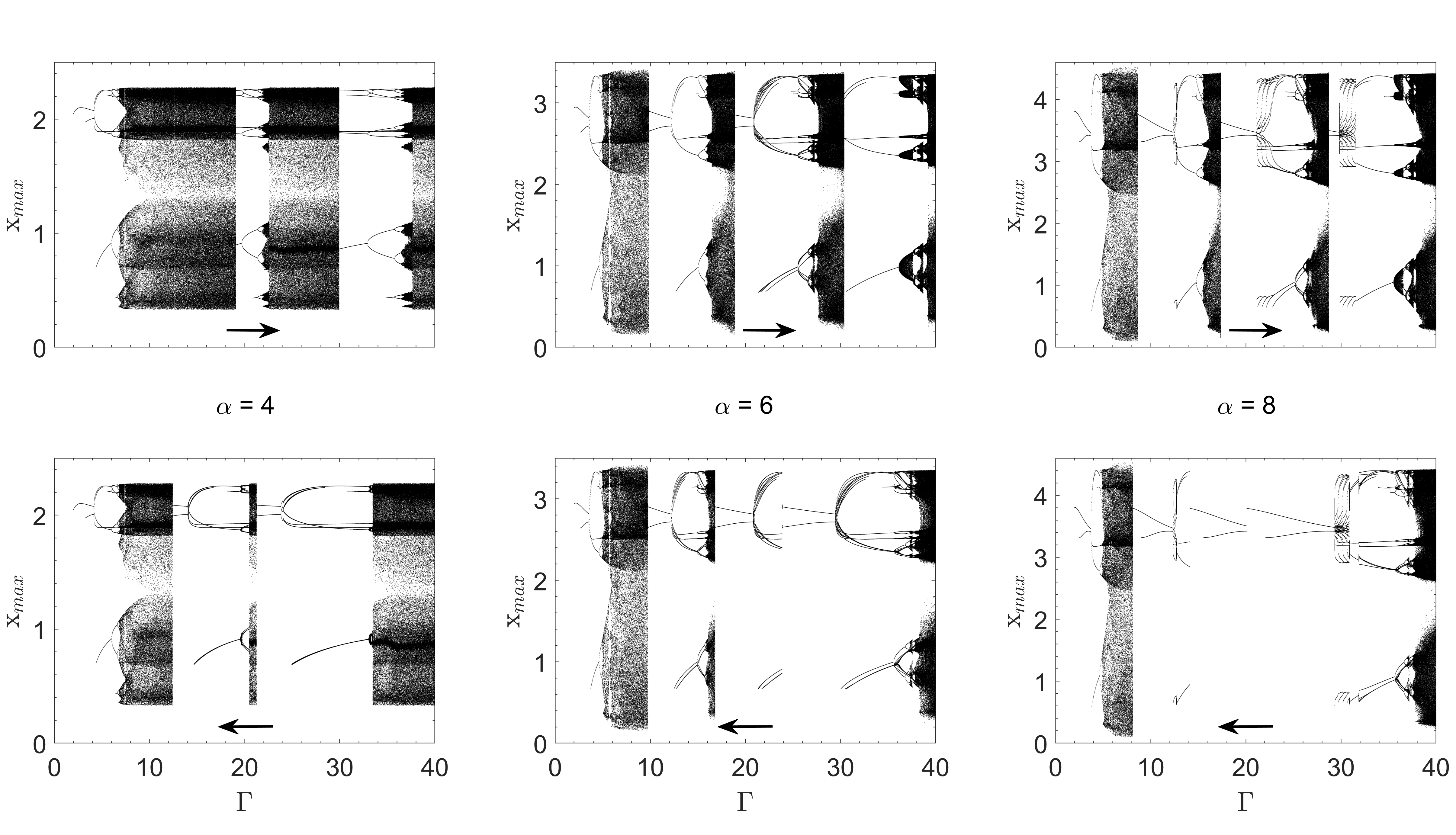}}
\caption{Bifurcation diagrams for different values
    $\alpha = 4$ (left column), $\alpha = 6$ (center column) and
    $\alpha = 8$ (right column). Peaks are recorded in a span of $200
    \Gamma$ after a transient of $2000 \Gamma$ was neglected. The
    initial condition for $\Gamma = 2$ was set to $x_{in}(t)=1$
    $\forall t\in(-\Gamma,0)$, then the previous solution was used as
    the initial condition as $\Gamma$ was swept up and then down. The
    top row corresponds to sweeping $\Gamma$ up whereas the bottom row
    correspond to sweeping $\Gamma$ down.} 
\label{figHysteresis}
\end{figure}

Figure \ref{figHysteresis} shows bifurcation diagrams where parameter
$\Gamma$ was swept up (top row) and down (bottom row) using the
previous solution as the initial condition (the first initial
condition function is $x_{in}(t)=1$ in the interval $(-\Gamma,
0)$). The three columns correspond to different values of
$\alpha$. Similar behaviour to that observed in
Fig.~\ref{figBifurcations} concerning that of period-doubling and
creation and destruction of branches is noticed, but in
Fig.~\ref{figHysteresis} periodic solutions appear for $\alpha = 4$
and $\Gamma > 8$ whereas in Fig.~\ref{figBifurcations} that region was
entirely chaotic, evidencing the multistability of the
system. Furthermore, the differences between the top and bottom row
diagrams suggest a sort of hysteresis loop since the diagrams vary as
$\Gamma$ is swept up or down. This hysteresis phenomena is further
hint of the multistability of the system as the initial conditions are
not the same when sweeping $\Gamma$ up to those when sweeping $\Gamma$
down. 

\begin{figure}
\centerline{\includegraphics[width=0.6\columnwidth]{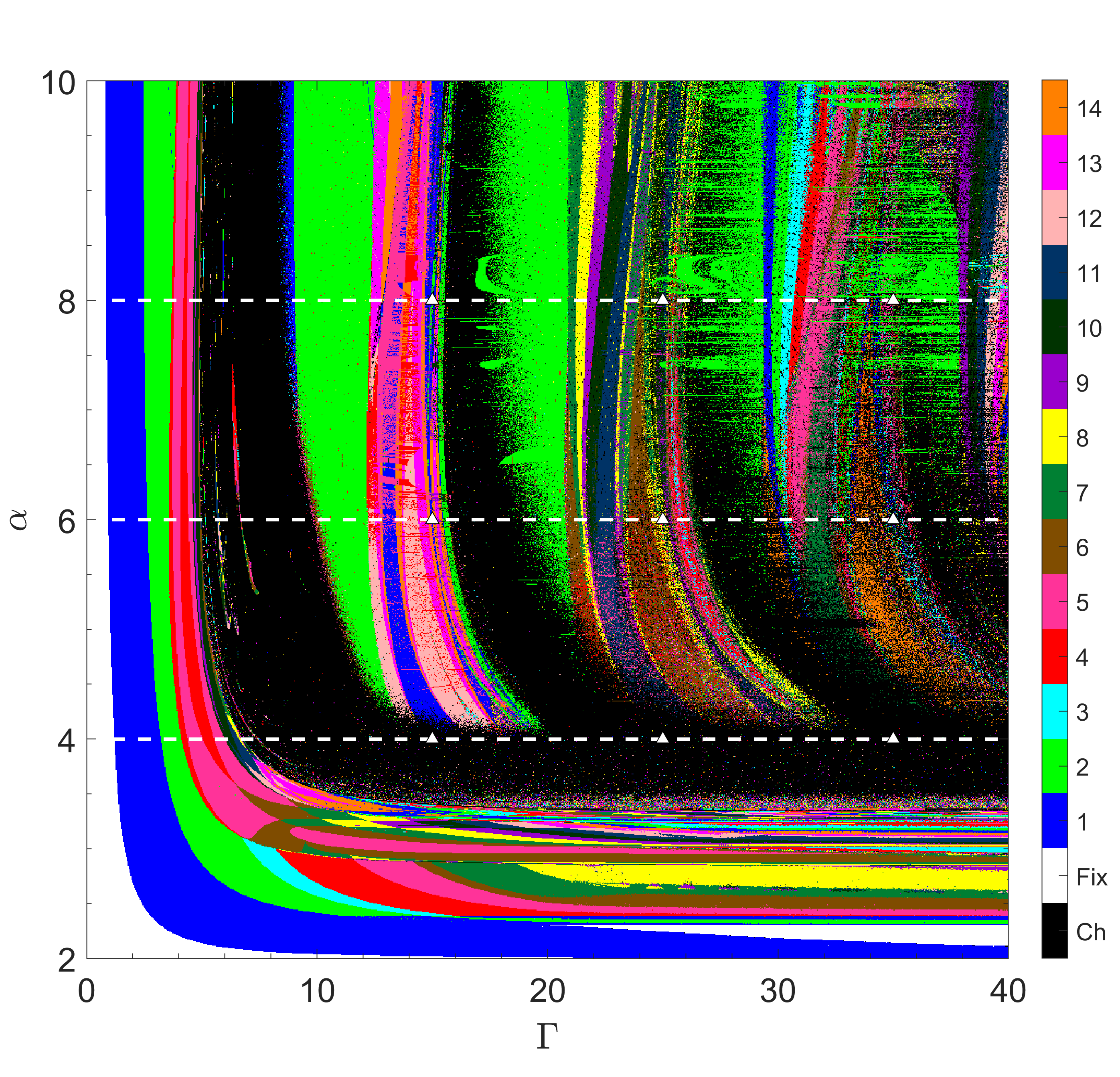}}
\caption{Isospike diagram obtained changing control parameters
  $\alpha$ and $\Gamma$. 1000x1000 parameter points are displayed. The
  number of peaks in a period is represented in colors according to
  the color bar shown in the right. Colors were recycled after 14 to
  represent more peaks per period. The initial conditions correspond
  to a constant value $x_{in}(t)=1$ $\forall t\in(-\Gamma,0)$ for each
  point of the diagram. To count the number of peaks a time of $200
  \Gamma$ was recorded after a transient time of $1000
  \Gamma$. Horizontal dashed lines correspond to the bifurcation
  diagrams of Fig.~\ref{figBifurcations}. }
\label{figParameterSpace}
\end{figure}

A more general picture of the global behavior can be obtained plotting
isospike diagrams in which a color scale indicates the number of peaks
in a given period of the variable or black in the case of chaotic
solutions \cite{freire2011stern,freire2013antiperiodic}.  Figure
\ref{figParameterSpace} shows an isospike diagram as parameters
$\alpha$ and $\Gamma$ were varied. As shown in this figure intricate
patterns arise in the parameter space, even in regions of high delay
$\Gamma$ where chaos is to be expected. Several regions of periodicity
are shown matching those seen in Fig.~\ref{figBifurcations} reveling
the complexity of the model as $\alpha$ and $\Gamma$ grow.  The
temporal series presented in Fig.~\ref{figTimeSeries} are also
coherent whit these figures revealing that a region consists of
similar series with variations due to creation and destruction of
peaks while different regions will have different series structures.
This patterns are examples of the intricacy and richness of delayed
systems.

\section{Impact of the multistability} 
\label{multistability}

The multistability in this system is ubiquitous in large regions of
the parameter space. To gain insight, we consider four initial
condition functions in the interval $ (-\Gamma,0)$: a non-null
constant value, a linear function, and two other periodic functions
\cite{amil2015organization} combining sinusoidal functions:
\begin{equation}
    x_{A}(t)=1 
    \label{eqA}
\end{equation}
\begin{equation}
    x_{B}(t)=0.7\frac{t}{\Gamma}+0.3
    \label{eqB}
\end{equation}
\begin{equation}
    x_{C}(t) = \frac{1}{4}\sin\Bigg(\frac{7\pi t}{\Gamma} + \phi\Bigg)\sin\Bigg(\frac{7\pi t}{2\Gamma} + \frac{\phi}{2}\Bigg) + x_{\mathrm{off}}
    \label{eqC}
\end{equation}
\begin{equation}
    x_{D}(t)=\frac{1}{40}\sin\bigg(\frac{7\pi t}{\Gamma} + \phi\bigg)\bigg[\sin\bigg(\frac{7\pi t}{2\Gamma} + \frac{\phi}{2}\bigg)+4\bigg] + x_{\mathrm{off}}.
    \label{eqD}
\end{equation}
Using these functions we characterize systematically the abundance of
coexisting solutions.

\begin{figure}
\centerline{\includegraphics[width=0.8\columnwidth]{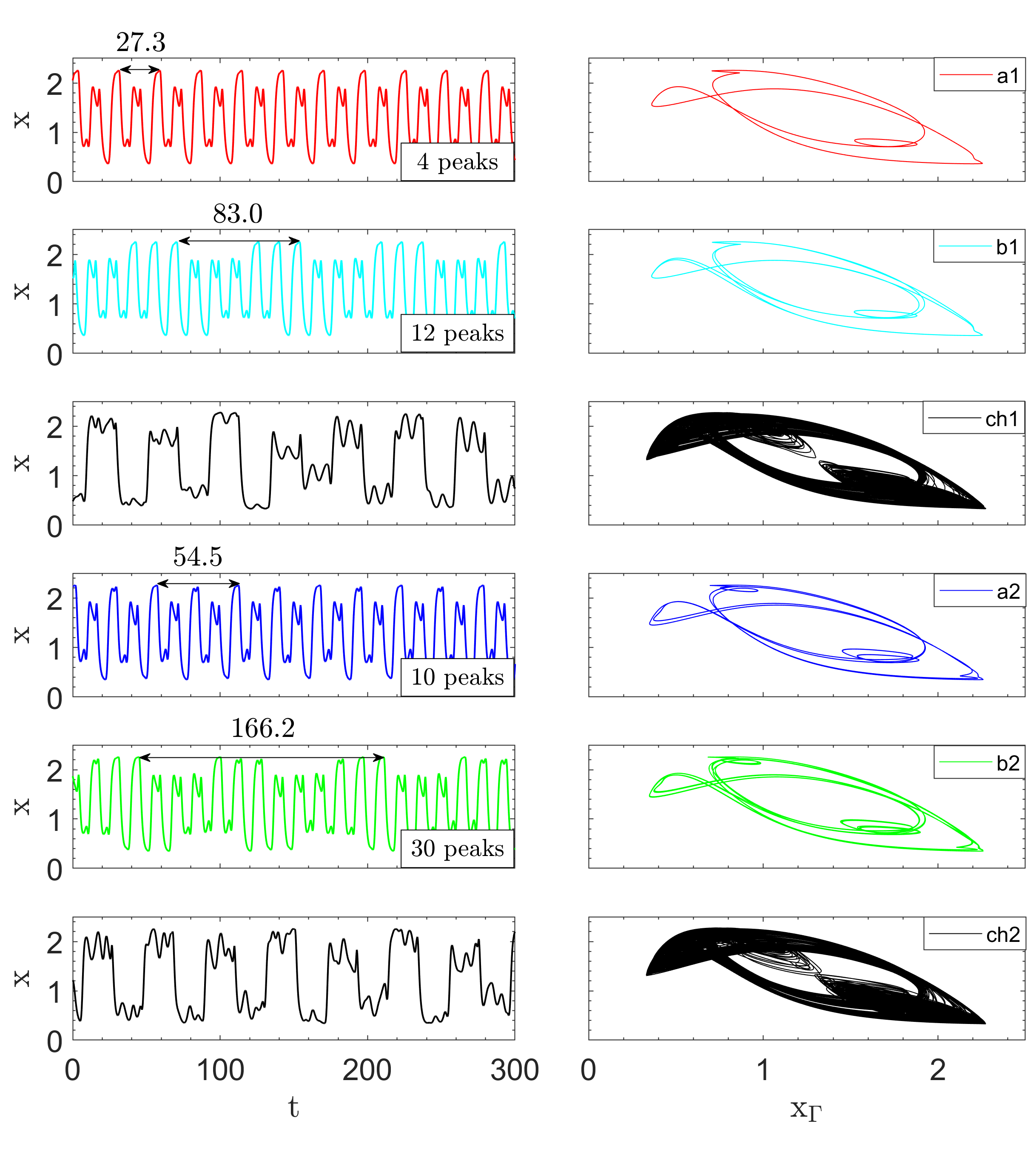}}
\caption{Coexisting solutions for different initial conditions in two
  fixed points of the parameter space. The temporal series and
  recurrence plot a1, b1 and ch1 correspond to the point $\Gamma =
  18$, $\alpha = 4$ and a2, b2 and ch2 correspond to $\Gamma = 20$,
  $\alpha = 4$. The initial condition function is given by
  Eq. \eqref{eqC} with control parameters $\phi = 0$ and
  $x_{\mathrm{off}} = 0.35$ for a1 and a2, $\phi = \pi$ and
  $x_{\mathrm{off}} = 0.3$ for b1 and b2 and $\phi = \pi$ and
  $x_{\mathrm{off}} = 0.4$ for ch1 and ch2. A transient time of $2000
  \Gamma$ was neglected.}
\label{figSeriesMulti}
\end{figure}

To illustrate the multistability, firstly, we selected two points
labelled 1 ($\Gamma_{1}=18$, $\alpha_{1}=4$) and 2 ($\Gamma_{2}=20$,
$\alpha_{2}=4$) and three different initial condition functions. These
functions were labelled \textbf{a}, \textbf{b}, \textbf{ch}, obtained
changing the parameters $\phi$ and $x_{\mathrm{off}}$ of
Eq. \eqref{eqC}. In Fig.~\ref{figSeriesMulti} we show six temporal
series (left column) and the respective recurrence plots (right
column) corresponding to the two mentioned points and the three
initial conditions. Both representation, temporal series and
recurrence plot, provide supplementary pictures of the dynamics.  In
all the cases, the solutions labeled \textbf{a} and \textbf{b} are
periodic while the corresponding to \textbf{ch} are chaotic. The
characteristics of the periodic solutions at each point depend on the
initial conditions, for example, in \textbf{a1} we observe a period of
27.3 and 4 peaks per period while in \textbf{a2} the period is 54.5
and there are 10 peaks per period.  Nevertheless, when comparing
panels a and b for both points, 1 and 2, the period and the number of
peaks is multiplied by 3 suggesting a similar bifurcation scenario.

\begin{figure}
\centerline{\includegraphics[width=0.85\columnwidth]{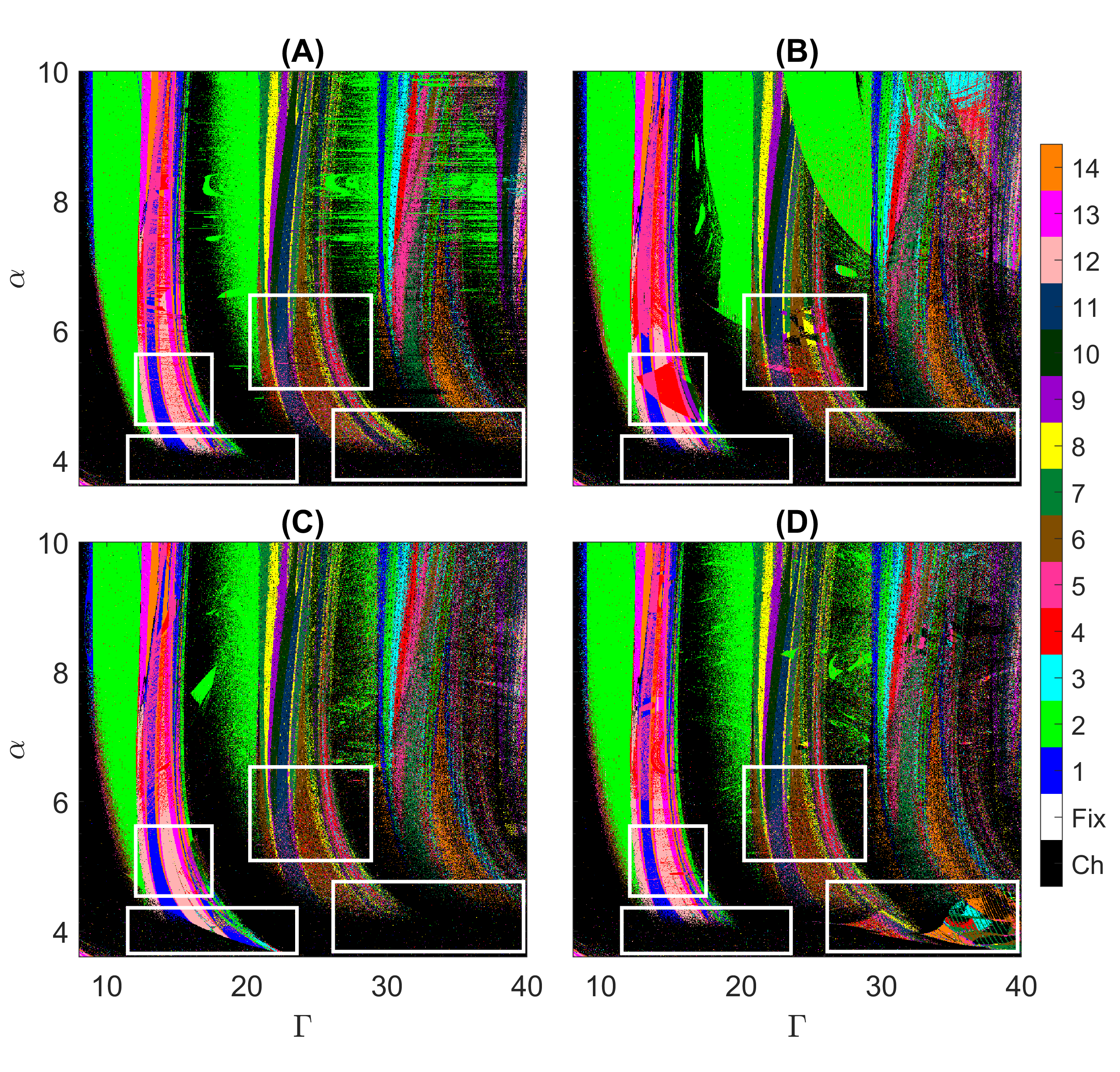}}
\caption{Isospike diagrams for different initial conditions
  functions. Each diagram displays 1000x1000 parameter points. The
  number of peaks in a period is represented in color. Colors were
  recycled after 14 to represent more peaks per period. The initial
  conditions are shown in Eq. \eqref{eqA} for \textbf{(A)},
  Eq. \eqref{eqB} for \textbf{(B)}, Eq. \eqref{eqC} with $\phi=0$ and
  $x_{\mathrm{off}}=0.3$ for \textbf{(C)} and Eq. \eqref{eqD} with
  $\phi=0$ and $x_{\mathrm{off}}=0.33$ for \textbf{(D)}. To count the
  number of peaks a time of $200 \Gamma$ was recorded after a
  transient time of $1000 \Gamma$.}
\label{figParameterSpaceMulti}
\end{figure}

Figure \ref{figParameterSpaceMulti} shows isospike diagrams
corresponding to the four initial condition functions,
Eqs.~\ref{eqA}-\ref{eqD}, in a large region of the parameter
space. Each point in these diagrams was obtained starting with the
same initial conditions function. In all the panels we appreciate
chaotic or periodic regions intermingled. Nevertheless, we observe
that although some regions are similar in all the panels there are
others that differs notably. To identify the regions which present
more variations we included white boxes in
Fig.~\ref{figParameterSpaceMulti}.  These variations demonstrate that
multistability is not evenly distributed in the parameter space. In
particular, for sufficiently low values of $\alpha$ or $\Gamma$, it
appear only one stable solution while increasing one or both
parameters leads in the majority of the cases to multistable behavior.

\begin{figure}
\centerline{\includegraphics[width=\columnwidth]{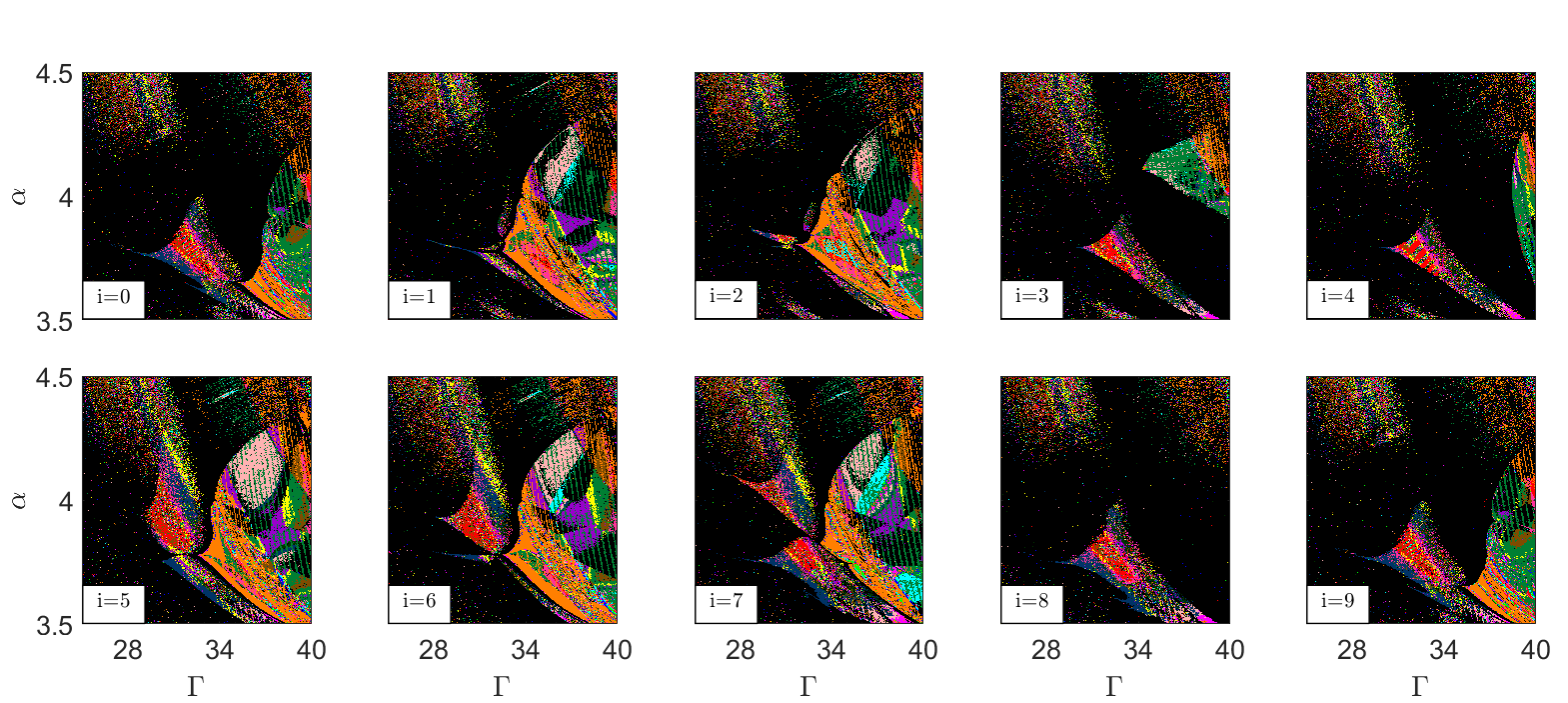}}
\caption{Parameter space diagrams for a family of initial conditions.
  The initial conditions were set to Eq. \eqref{eqD} with
  $x_{\mathrm{off}} = 0.35$ and $\phi = \frac{4\pi i}{10}$ with $i =
  0,1,... \, 9$. Each diagram displays 500x500 parameter points. To
  count the number of peaks, a time of $200 \Gamma$ was recorded after
  a transient time of $2000 \Gamma$.}
\label{fig10Params2540}
\end{figure}

\begin{figure}
\centerline{\includegraphics[width=\columnwidth]{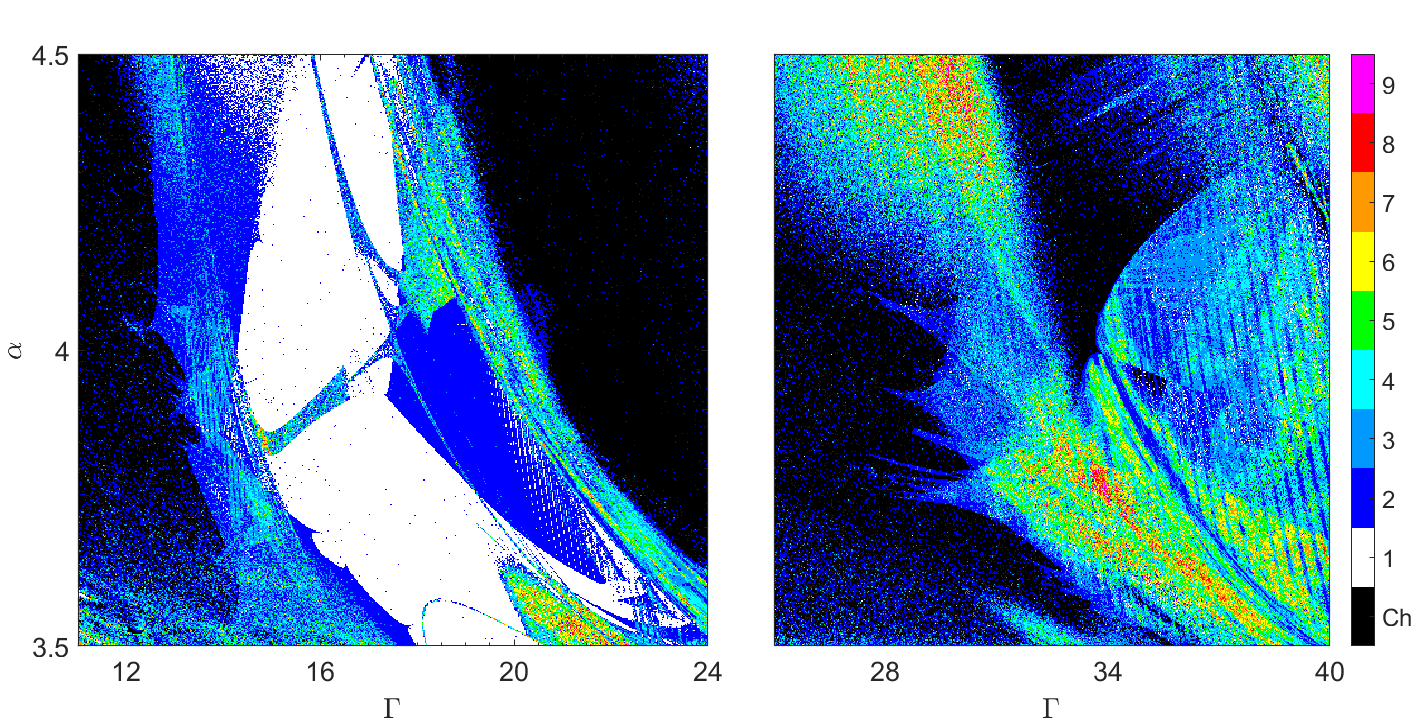}}
\caption{Number of distinct solutions as the initial conditions were
  varied for a region of the parameter space. The initial conditions
  were set to Eq. \eqref{eqC} for the left diagram and Eq. \eqref{eqD}
  for the right diagram with $x_{\mathrm{off}} = 0.35$ and $\phi =
  \frac{4\pi i}{10}$ with $i = 0,1,... \, 9$ and the number of peaks
  was recorded for each diagram. The number of different solutions is
  represented in color for each point of the parameter space. Each
  diagram displays 500x500 parameter points.  To count the number of
  peaks a time of $200 \Gamma$ was recorded after a transient time of
  $2000 \Gamma$.}
\label{figNumSolutions}
\end{figure}

To further quantify multistability, we selected a family of functions,
given by Eq.~\ref{eqD}, and changed systematically the parameter
values. Then, the solutions obtained were analyzed for different
control parameter values.  Figure \ref{fig10Params2540} shows the
changes in peaks count for a smaller region of the parameter space as
the initial conditions are changed. The initial conditions for each
diagram correspond to the Eq. \eqref{eqD} with $x_{\mathrm{off}} =
0.35$ and $\phi = \frac{4\pi i}{10}$ for $i = 0,1,... \, 9$. Small
variations between the diagrams indicate the coexistence of multiple
solutions for different initial conditions, even when only the phase
is changed. To further assess this observation, the right diagram of
Fig.~\ref{figNumSolutions} shows the number of distinct solutions that
appear in Fig.~\ref{fig10Params2540} for every point of the parameter
space, revealing a region where no periodic solutions were observed
(black) and regions where multiple solutions coexist for different
initial conditions functions (color). The left diagram of
Fig.~\ref{figNumSolutions} shows the same analysis for a different
region of the parameter space and a the family of initial conditions
functions given by \eqref{eqC} with $x_{\mathrm{off}} = 0.35$ and
$\phi = \frac{4\pi i}{10}$ for $i = 0,1,... \, 9$. For this condition,
a region without multistability (white) clearly appears.

\begin{figure}
\centerline{\includegraphics[width=\columnwidth]{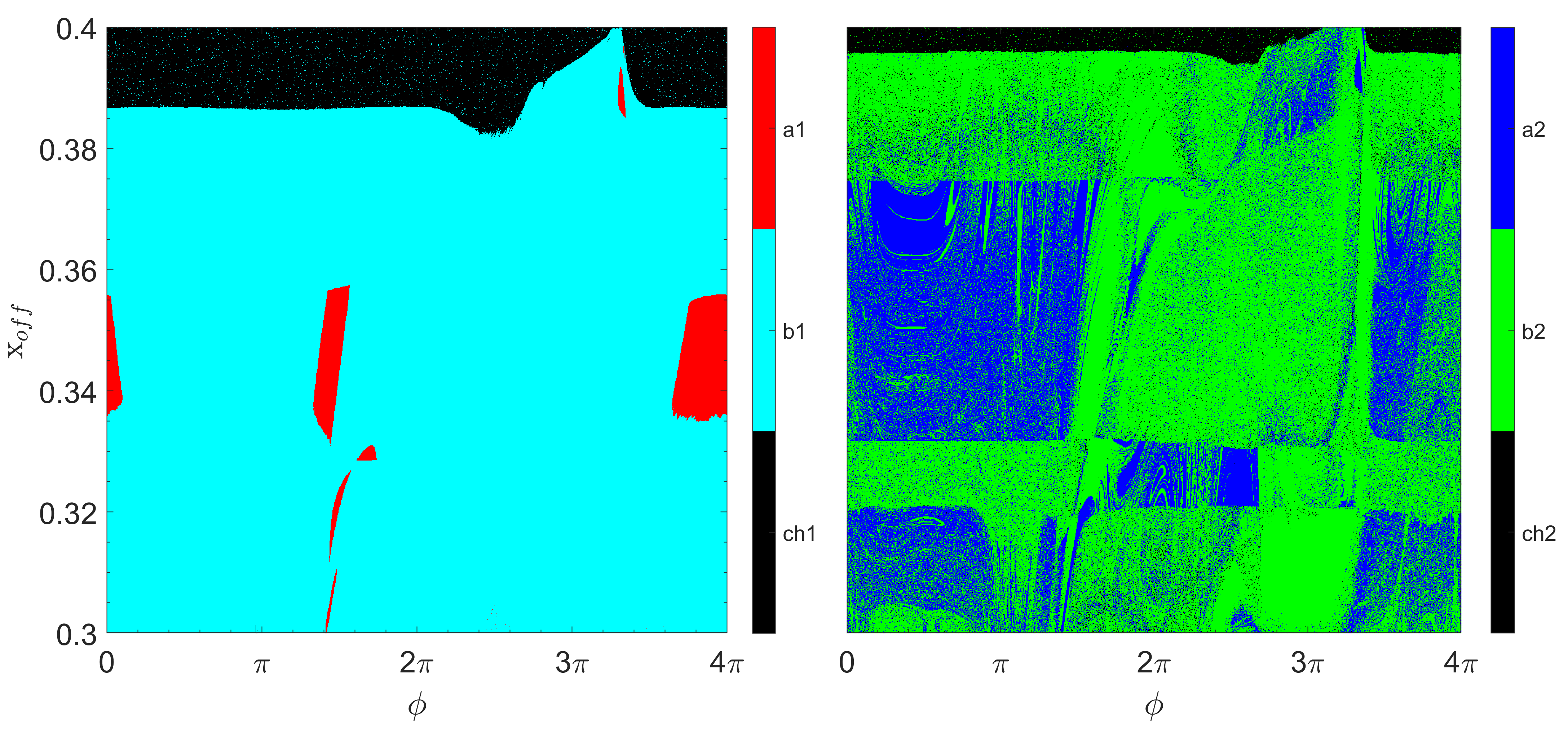}}
\caption{Map of coexisting solutions in a fixed point of the parameter
  space as the initial conditions are changed. Each diagram displays
  1000x1000 parameter points. The left diagram corresponds to the
  point $\alpha = 4$, $\Gamma = 18$ and the diagram to the right to
  the point $\alpha = 4$, $\Gamma = 20$. The initial conditions were
  set to Eq. \eqref{eqC} as parameters $\phi$ and $x_{\mathrm{off}} $
  were varied. A transient time of $2000 \Gamma$ was neglected.}
\label{figMultistability}
\end{figure}

A distinctive region of two solutions is shown in the left diagram of
Fig.~\ref{figNumSolutions} for the 10 initial conditions functions
selected. To continue with the study of multistability in this region
two points $\Gamma = 18$, $\alpha =4$ and $\Gamma = 20$, $\alpha =4$
were selected and a map of coexisting solutions was made as the
initial conditions vary inside the same family of functions. Figure
\ref{figMultistability} shows the different solutions that arise for
the points $\Gamma = 18$, $\alpha =4$ (left diagram) and $\Gamma =
20$, $\alpha =4$ (right diagram), as initial conditions are
changed. Initial conditions were set to Eq. \eqref{eqC} and parameters
$\phi$ and $x_\mathrm{off}$ were varied finding two periodic solutions
and a region of chaos for both points. Time series are shown in
Fig~\ref{figSeriesMulti}. Similar patterns are observed between both
diagrams, primarily the fact that regions of \textbf{a1} in the left
diagram correspond to regions of \textbf{a2} in the right diagram and
some structure remains on the frontier with the chaotic solutions.

\section{Conclusion}

The Mackey-Glass delayed model was studied in this work using a
previously validated numerical scheme. By means of temporal series,
bifurcation and isospike diagrams we explored the parameter space
expressed in terms of the dimensionless production rate and decay. In
general terms, the complexity increases as increasing the delay. The
presence of tongues of stability in the isospike diagrams is
characteristics of this system.

Varying the initial condition function we found the coexistence of
several solutions, i.e. multistability, in a broad range of values of
space parameters.  Moreover, periodic solutions consisting of maximums
of several amplitudes but in different order are abundant in various
of these regions.  We characterized the coexisting
  solutions, either periodic or chaotic, in the parameter
  space. Selecting representative families of initial condition
  function we quantitatively evaluated the impact of the
  multistability. We also showed the existence of hysteresis loops by
  sweeping up and down the delay parameter in the bifurcation
  diagrams. Undoubtedly, the range of possible applications of
delayed systems will continue to enlarge in the near future.

\section*{Acknowledgments}
The authors would like to thank the Uruguayan institutions Programa de
Desarrollo de las Ciencias Básicas (MEC-Udelar, Uruguay) and Comisión
Sectorial de Investigación Científica (Udelar, Uruguay) for the grant
Física Nolineal (ID 722) Programa Grupos I+D. The numerical
experiments presented here were performed at the ClusterUY (site:
https://cluster.uy)

\bibliographystyle{unsrt}
\bibliography{ref}

\begin{thebibliography}{10}

\bibitem{mackey1977oscillation}
Michael~C Mackey and Leon Glass.
\newblock Oscillation and chaos in physiological control systems.
\newblock {\em Science}, 197(4300):287--289, 1977.

\bibitem{belair1995dynamical}
Jacques B{\'e}lair.
\newblock {\em Dynamical Disease: Mathematical Analysis of Human Illness;[the
  Papers are Based on a NATO Advanced Research Workshop Held in Mont Tremblant,
  Qu{\'e}bec, Canada in February 1994]}.
\newblock AIP Press, 1995.

\bibitem{biswas2018time}
Debabrata Biswas and Tanmoy Banerjee.
\newblock {\em Time-Delayed Chaotic Dynamical Systems}.
\newblock Springer, 2018.

\bibitem{hale2013introduction}
Jack~K Hale and Sjoerd M~Verduyn Lunel.
\newblock {\em Introduction to functional differential equations}, volume~99.
\newblock Springer Science \& Business Media, 2013.

\bibitem{junges2012intricate}
Leandro Junges and Jason~AC Gallas.
\newblock Intricate routes to chaos in the mackey--glass delayed feedback
  system.
\newblock {\em Physics Letters A}, 376(18):2109--–2116, 2012.

\bibitem{losson1993solution}
Jerome Losson, Michael~C Mackey, and Andre Longtin.
\newblock Solution multistability in first-order nonlinear differential delay
  equations.
\newblock {\em Chaos: An Interdisciplinary Journal of Nonlinear Science},
  3(2):167--176, 1993.

\bibitem{mensour1995controlling}
Boualem Mensour and Andr{\'e} Longtin.
\newblock Controlling chaos to store information in delay-differential
  equations.
\newblock {\em Physics Letters A}, 205(1):18--24, 1995.

\bibitem{lim1998experimental}
Tong~Kun Lim, Keumcheol Kwak, and Mijeong Yun.
\newblock An experimental study of storing information in a controlled chaotic
  system with time-delayed feedback.
\newblock {\em Physics Letters A}, 240(6):287--294, 1998.

\bibitem{zhou2007isochronal}
Brian~B Zhou and Rajarshi Roy.
\newblock Isochronal synchrony and bidirectional communication with
  delay-coupled nonlinear oscillators.
\newblock {\em Physical Review E}, 75(2):026205, 2007.

\bibitem{pyragas1998synchronization}
Kestutis Pyragas.
\newblock Synchronization of coupled time-delay systems: Analytical
  estimations.
\newblock {\em Physical Review E}, 58(3):3067, 1998.

\bibitem{kim2006synchronization}
Min-Young Kim, Christopher Sramek, Atsushi Uchida, and Rajarshi Roy.
\newblock Synchronization of unidirectionally coupled mackey-glass analog
  circuits with frequency bandwidth limitations.
\newblock {\em Physical Review E}, 74(1):016211, 2006.

\bibitem{shahverdiev2006chaos}
EM~Shahverdiev, RA~Nuriev, RH~Hashimov, and KA~Shore.
\newblock Chaos synchronization between the mackey--glass systems with multiple
  time delays.
\newblock {\em Chaos, Solitons \& Fractals}, 29(4):854--861, 2006.

\bibitem{namajunas1995electronic}
A~Namaj{\=u}nas, K~Pyragas, and A~Tama{\v{s}}evi{\v{c}}ius.
\newblock An electronic analog of the mackey-glass system.
\newblock {\em Physics Letters A}, 201(1):42--46, 1995.

\bibitem{amil2015exact}
Pablo Amil, Cecilia Cabeza, and Arturo~C Marti.
\newblock Exact discrete-time implementation of the mackey--glass delayed
  model.
\newblock {\em IEEE Transactions on Circuits and Systems II: Express Briefs},
  62(7):681--685, 2015.

\bibitem{amil2015organization}
Pablo Amil, Cecilia Cabeza, Cristina Masoller, and Arturo~C Mart{\'\i}.
\newblock Organization and identification of solutions in the time-delayed
  mackey-glass model.
\newblock {\em Chaos: An Interdisciplinary Journal of Nonlinear Science},
  25(4):043112, 2015.

\bibitem{smith2011introduction}
Hal~L Smith.
\newblock {\em An introduction to delay differential equations with
  applications to the life sciences}, volume~57.
\newblock Springer New York, 2011.

\bibitem{bellen2013numerical}
Alfredo Bellen and Marino Zennaro.
\newblock {\em Numerical methods for delay differential equations}.
\newblock Oxford university press, 2013.

\bibitem{freire2011stern}
Joana~G Freire and Jason~AC Gallas.
\newblock Stern--brocot trees in cascades of mixed-mode oscillations and
  canards in the extended bonhoeffer--van der pol and the fitzhugh--nagumo
  models of excitable systems.
\newblock {\em Physics Letters A}, 375(7):1097--1103, 2011.

\bibitem{freire2013antiperiodic}
Joana~G Freire, Cecilia Cabeza, Arturo Marti, Thorsten P{\"o}schel, and
  Jason~AC Gallas.
\newblock Antiperiodic oscillations.
\newblock {\em Scientific Reports}, 3, 2013.

\end{thebibliography}


\end{document}